\documentclass[12pt]{article}
\usepackage{amsmath}
\usepackage{amssymb}
\usepackage[titletoc]{appendix}
\begin{document}
\vskip 2cm
\begin{center}
{\sf {\Large  Novel symmetries in an interacting $\mathcal{N }= 2$ supersymmetric quantum mechanical model}}

\vskip 3.0cm

{\sf S. Krishna$^{(a)}$,  D. Shukla$^{(a)}$, R. P. Malik$^{(a,b)}$}\\
$^{(a)}$ {\it Physics Department, Centre of Advanced Studies,}\\
{\it Banaras Hindu University, Varanasi - 221 005, (U.P.), India}\\

$^{(b)}$ {\it DST Centre for Interdisciplinary Mathematical Sciences,}\\
{\it Faculty of Science, Banaras Hindu University, Varanasi - 221 005, India}\\
{\small {\sf {e-mails: skrishna.bhu@gmail.com;  dheerajkumarshukla@gmail.com; rpmalik1995@gmail.com}}}

\end{center}

\vskip 2cm

\noindent
{\bf Abstract:} 
We demonstrate the existence of a set of novel discrete symmetry transformations in the case of  
an interacting $\mathcal{N} = 2$ supersymmetric quantum mechanical model of a system of an electron moving
on a sphere in the background of a magnetic monopole  
and establish its interpretation in the language of differential geometry. 
These discrete symmetries are, over and above, the usual {\it three} continuous 
symmetries of the theory which {\it together} provide the physical realizations  of 
the de Rham cohomological operators of differential geometry. We derive the nilpotent 
$\mathcal{N} = 2$ SUSY transformations by exploiting our idea of supervariable approach 
and provide geometrical meaning to these transformations in the language of Grassmannian  
translational generators on a (1, 2)-dimensional supermanifold on which our $\mathcal{N} 
= 2$ SUSY quantum mechanical model is generalized. We express the conserved supercharges and the
invariance of the Lagrangian in terms of the supervariables (obtained after the imposition of the 
SUSY invariant restrictions) and provide the geometrical meaning to (i) the nilpotency property 
of the $\mathcal{N} = 2$ supercharges, and (ii) the SUSY invariance of the Lagrangian of our  
 $\mathcal{N} = 2$ SUSY theory.



\newpage
\section{Introduction}

It is a well-known fact that three (out of four) fundamental interactions of nature are governed
by the gauge theories which are endowed with the first-class constraints in the language of Dirac's
prescription for the classification scheme. These theories are characterized by the existence of 
local gauge symmetries 
which are generated by the above first-class constraints.
There is a class of gauge theories that respect the dual-gauge symmetry transformations in addition
to the above cited local gauge  symmetry transformations. Such gauge theories provide the 
physical models for the Hodge
theory within the framework of Becchi-Rouet-Stora-Tyutin (BRST) formalism where the local gauge symmetries
are traded with the nilpotent (anti-)BRST symmetries and the dual-gauge symmetry transformations are elevated
to the (anti-) co-BRST symmetry transformations at the {\it quantum} level. In an earlier  article 
(see, e.g. [1] and references therein), we
have shown that such examples of gauge theories are  the Abelian $p$-form ($p = 1, 2, 3 $) gauge theories that
are described within the framework of BRST formalism in $D = 2 p$ dimensions of spacetime.

In a recent set of papers [2-5], we have established that the $\mathcal{N} = 2$ 
supersymmetric (SUSY) quantum mechanical models (QMMs) {\it also} provide a set of tractable physical 
examples of Hodge theory because their continuous symmetries (and corresponding conserved 
charges) provide the physical realizations of the de Rham cohomological operators\footnote{On a compact
 manifold without a boundary, there exists a set of three differential operators ($d , \delta, \Delta$)
which are called as the de Rham cohomological  operators  of differential geometry.
The (co-) exterior derivatives $(\delta)d$ are nilpotent of order two and their anticommutator
defines the Laplacian operator $\Delta$. They follow the algebra: $d^2 = \delta^2 = 0, \Delta
= (d + \delta)^2 = \{ d, \delta \}, [\Delta, d]= [\Delta, \delta] = 0$. The (co-)exterior derivatives
are connected with each-other by the relationship: $\delta = \pm * d *$ where the $(*)$ stands
for the Hodge duality operation on the given compact manifold. The Laplacian operator behaves like
the Casimir operator (see, e.g. [6-10]) for this algebra.} 
of differential geometry [6-10] and discrete symmetry transformations correspond to the 
physical analogue of the Hodge duality operation. We have also established the exact similarities 
between the Hodge algebra obeyed by the cohomological operators and algebra respected by the 
conserved SUSY charges and the Hamiltonian of the theory (which is nothing but one of the 
simplest $\mathcal{N} = 2$ SUSY algebra (i.e. $sl(1/1)$) without any central extension).

Such studies are important  {\it physically} because exploiting the inputs 
from these  kind of studies, we have been able to 
establish  [11] that the two (1 + 1)-dimensional (2D) {\it free}  (non-)Abelian 1-form gauge theories 
(without any interaction with matter fields) are the {\it new} models for the topological field 
theories (TFTs) which  capture some salient features of Witten-type TFTs and a few key features of 
Schwarz-type TFTs (see, e.g. [11-13] for details). We have {\it also} shown that the {\it above} free 
gauge field theoretic models and the 2D $U(1)$ gauge theory interacting with the Dirac fields 
[14,15] are the {\it perfect} models for the Hodge  theory within the framework of  
BRST formalism. Such a set of Hodge theoretic models have
 {\it also} been proven in the description 
of the 2D modified versions of the Abelian 1-form anomalous gauge theory as well as Proca theory (where  the gauge and 
matter fields are present [16,17] along with the mass term for the gauge fields).

In a very recent paper [5], we have established that the free version of the widely-studied
{\it interacting} $\mathcal{N} = 2$ SUSY QMM of a charged particle (i.e. an electron) constrained to move 
on a sphere, in the background of a magnetic monopole, is a model for the Hodge theory. 
The purpose of our present paper is to demonstrate that, in addition to the free case, 
the {\it interacting} $\mathcal{N} = 2$  SUSY QMM of the above system {\it also} provides a tractable 
physical example of Hodge theory where the conserved Noether charges  of this theory follow  exactly the same 
algebraic structure as that of the de Rham cohomological operators of differential geometry. 
Furthermore, we apply the theoretical arsenal of supervariable approach\footnote{We christen
our present approach as the supervariable approach because when we set the Grassmannian
variables equal to zero in a supervariable (defined on the (1, 2)-dimensional
supermanifold), we obtain an ordinary variable which is a function of $t$ only in the realm
of SUSY quantum mechanics.} [18-21] to derive the nilpotent 
$\mathcal{N} = 2$ SUSY transformations of this theory and provide the geometrical 
interpretations for the nilpotency 
property of the  $\mathcal{N} = 2$ SUSY transformations (and their {\it generators})
as well as the SUSY invariance of the Lagrangian of our theory.

In the context of the statements made on the  method of supervariable approach
to derive the $\mathcal{N} = 2$ SUSY symmetries, we would like to mention that we 
have to concentrate on the (anti-)chiral supervariables so that we could capture
{\it only} the nilpotency property of the SUSY symmetries (and avoid the property of absolute anticommutativity).
In our present endeavor (and its predecessor [5]) we have accomplished this goal and we 
have provided the geometrical basis for the nilpotency and SUSY invariance of the Lagrangian for 
our present theory. The choice of the (anti-)chiral supervariables should be contrasted with the superfield
approach to BRST formalism (see, e.g. [22-27]) where the superfields\footnote{When we set
the Grassmannian variables equal to zero in a superfield (defined on a (D, 2)-dimensional
supermanifold), we obtain an ordinary field which is a function of the $D$-dimensional spacetime 
ordinary coordinates. This is why, we christen the approach, adopted in the realm of BRST formalism,
as the superfield approach because it is applied to the description of an ordinary $D$-dimensional gauge field theory.}, 
defined on the (D, 2)-dimensional
supermanifold, are expanded along {\it all} the Grassmannian directions of the supermanifold 
so that one could capture
the nilpotency as well as the anticommutativity properties of the (anti-)BRST symmetries for a given
$D$-dimensional gauge theory which is generalized onto the above supermanifold.

The main motivating factors  behind our present endeavor are as follows. 
First and foremost, it is important for us to prove that the 
{\it interacting} {$\mathcal{N} = 2$} SUSY QMM of the motion of an electron on a sphere in the background 
of a magnetic monopole is {\it also} a model for the Hodge theory (as has been shown by us that its 
{\it free} version is a tractable model for the Hodge theory [5]). Second, the physical realization of 
the abstract mathematical de Rham cohomological operators in the language of discrete and continuous symmetry transformations of a physical theory is {\it interesting } in its {\it own right}. Third, the goal of proving 
various SUSY models (e.g. $\mathcal{N} = 2, 4, 8...$) to be the models for Hodge 
theory is theoretically important as it might turn out to
be useful in the study of {\it such} SUSY gauge theories (in various dimensions of spacetime)  
which are important, at the moment, because of their connection  with the superstring theories. 
Finally,  our present endeavor (and its 
predecessor [5]), are our modest steps towards  our main goal of proving the interacting $\mathcal{N} = 4$
SUSY QMM to provide a tractable SUSY model for the Hodge theory (because the generalization 
of our present interacting $\mathcal{N} = 2$ SUSY QMM to its counterpart $\mathcal{N} = 4$ SUSY
quantum theory has already been discussed in the literature [28]).

The contents of our present investigation are organized as follows. In Sec. 2, we recapitulate 
the bare essentials of the nilpotent $\mathcal{N} = 2$ SUSY transformations and show that their 
anticommutator generates a bosonic symmetry in the theory. We compute the conserved charges by 
exploiting the basic concepts of Noether's theorem. Our Sec. 3 is devoted to the discussion of 
a set of novel discrete symmetry transformations which turn out to be responsible for establishing a useful 
connection between the two nilpotent  $\mathcal{N} = 2$ SUSY transformations. In Sec. 4, we derive the algebraic 
structure of the symmetry transformations in their operator form and show the existence of one of 
the simplest form of $\mathcal{N} = 2$ SUSY algebra amongst the conserved charges. Our Sec. 5 deals 
with the connection between $\mathcal{N} = 2$ SUSY algebra and the de Rham cohomological operators 
of differential geometry  thereby providing the proof that our present model is a SUSY quantum 
mechanical example of a Hodge theory. Sec. 6 deals with the derivation of $\mathcal{N} = 2$ SUSY 
transformations by exploiting the supervariable approach. Finally, we make some concluding remarks 
in our Sec. 7 and point out a few future directions for further investigation.

In our Appendix A, for the readers' convenience, we mention a few key points that are
 connected with the superspace formalism  where the  
(anti-)chiral supervariables are used [29]. 
  Our Appendices  B, C and D are devoted to clarify some of the  expressions/equations  
that have been used in the main body of our  text.

\section{Preliminaries: Lagrangian formulation }

Let us begin with the following  Lagrangian for the $\mathcal{N} = 2$ SUSY quantum 
mechanical model of the motion an electron   on  a  sphere in the background of the Dirac's
magnetic monopole based on the $CP^{(1)}$-model approach   (see, e.g. [29] for details)       
\begin{eqnarray}
L &=& 2\, (D_t {\bar z}) \cdot (D_t z) + \frac{i}{2}\, \left[\bar\psi \cdot (D_t\psi)
- (D_t {\bar\psi})\cdot \psi \right] - 2\, g\, a, 
\end{eqnarray} 
where $t$ is the evolution parameter in the theory and $\partial_t = {\partial}/{\partial t}$. 
Here the dot product between $\bar z$ and $z$ is taken to be $\bar z \cdot z = \Sigma_{i = 1}^2 |z_i|^2$ 
which clearly demonstrates  that 
\begin{eqnarray}
z  = \begin{pmatrix} z_1 \\ z_2 \end{pmatrix}  , \qquad \bar z =  \begin{pmatrix}  z_1^* & z_2^* \end{pmatrix} 
\Longrightarrow \bar z\cdot z = |z_1|^2 + |z_2|^2.
\end{eqnarray}
Similarly, other dot products, defined in the Lagrangian (1), should be taken into account.
We also have the ``covariant" derivatives $D_t z = (\partial_t - i\,a )\,z,\, \ D_t \bar z= (\partial_t 
+ i\,a )\,\bar z, \, D_t\psi=  (\partial_t - i\,a )\,\psi,\,D_t \bar\psi = (\partial_t + i\,a )
\,\bar\psi$ where $a$ is the ``gauge" variable
 and the complex variables $z$ and $\bar z$ are bosonic 
in nature while their superpartners $\psi$ and $\bar \psi$ are fermionic (i.e. $\psi^2 = \bar\psi^2 = 0, 
\psi\,\bar\psi + \bar\psi\,\psi = 0$) at the classical level\footnote{We lay empashis on the fact that the absolute anticommutativity 
($\psi\,\bar\psi + \bar\psi\,\psi = 0$) is true only at the {\it classical} level. As is evident from equation (9) (see, below),
at the {\it quantum} level, $\psi$ and $\bar\psi$ would {\it not} absolutely anticommute because they are canonically
 conjugate to each-other.}. In the Lagrangian (1), the parameter $g$ 
stands for the charge on the monopole which interacts with the electron via gauge variable $a$ through the coupling 
($- 2\,g\, a$). This  is the last term in the Lagrangian (1).   The charge on the electron 
has been set equal to ($- 1$) and its mass has been taken to be ($+ 1$).

The above Lagrangian respects the following infinitesimal,
 continuous and nilpotent ($s_1^2 = s_2^2 = 0$) 
$\mathcal{N} = 2$ SUSY transformations ($s_1, \, s_2$): 
\begin{eqnarray*}
&& s_1\, z = \frac{\psi}{\sqrt{2}}, \qquad s_1\, \psi =0, \qquad s_1\, 
\bar\psi = \frac{2\,i\, D_t\, \bar z} {\sqrt{2}}, \qquad s_1 \,\bar z = 0,
\nonumber\\ && s_1\,(D_t\,z) = \frac{D_t\, \psi}{\sqrt{2}},\qquad \qquad s_1\,(D_t\,\bar z) = 0, 
\qquad\qquad s_1 \,a = 0,\nonumber\\
\end{eqnarray*} 
\begin{eqnarray}
&& s_2\, \bar z =  \frac{\bar\psi}{\sqrt{2}}, \qquad s_2\, \bar\psi = 0, \qquad s_2\, \psi 
= \frac{2 \,i\, D_t\, z}{\sqrt{2}}, 
\qquad s_2 \, z = 0, \nonumber\\ 
&&  s_2\,(D_t\, \bar z) = \frac{ D_t\, \bar\psi}{\sqrt{2}}, \qquad\qquad s_2\,(D_t\, z) = 0,\qquad \qquad s_2 \,a = 0,
\end{eqnarray}
because the Lagrangian (1) transforms to the total time  derivatives  as   
\begin{eqnarray}
s_1\, L = \frac{d}{dt}\,\left[\frac{(D_t\,\bar z)\cdot\psi}{\sqrt{2}} \right], \qquad\qquad s_2\, L 
= \frac{d}{dt}\,\,\left[\frac{ \bar\psi \cdot (D_t\, z)}{\sqrt{2}}\right].    
\end{eqnarray}
The above expression demonstrates that the $\mathcal{N} = 2$ SUSY transformations $s_1$ and $s_2$ are the 
{\it symmetry} transformations for the action integral  $S = \int dt\, L$.  It is good to mention that
the ``gauge" variable, defined in the following fashion\footnote{Actual superspace formalism [29] 
yields the expression  for this ``gauge" variable (see, Appendix A) as:
$ a = {-} \,\left([ i\,(\bar{z}\cdot \dot{z}
 - \dot{\bar{z}} \cdot z) + (\bar{\psi}\cdot \psi)]/ [{2\,\bar{z} \cdot z}] \right).$ However, 
the substitution of the constraint $\bar{z} \cdot z = 1$ leads to the expression for $a$ 
as quoted in (5) which avoids the presence of the variables in the denominatorthe as well as singularity in our present theory.} 
\begin{eqnarray}
a = - \frac{i}{2}\, (\bar z \cdot\,\dot z - \dot{\bar z} \cdot z) - \frac{1}{2}\, (\bar\psi \cdot\psi),
\end{eqnarray}
remains invariant\footnote{The explicit application of $s_1$ and $s_2$ on $a$ yields the results: $s_1\, a = \frac{d}{dt} \,[- \frac{i}{2\sqrt 2}\,(\bar{z}\cdot \psi)] + \frac{a}{\sqrt 2}\,(\bar{z}\cdot \psi)$ and $s_2\, a = \frac{d}{dt}\, [+ \frac{i}{2 \sqrt 2}\,(\bar{\psi}\cdot z)] + \frac{a}{\sqrt 2}\,(\bar{\psi}\cdot z) $ which, finally, imply that $s_1\, a = s_2 \,a = 0$ 
due to the constraint  conditions: $\bar{z}\cdot \psi = 0$ and $\bar{\psi}\cdot z = 0$ which define the supersymmetrized version of the sphere ($\bar z \cdot z = 1$).} ($s_1\, a = s_2\, a = 0$) 
under the $\mathcal{N} = 2$ SUSY transformations $s_1$ and $s_2$ due to the supersphere defined by the following constraints  
(see, e.g. [29])
\begin{eqnarray}
\bar z \cdot z - 1 = 0, \qquad \qquad \bar z \cdot \psi = 0, \qquad\qquad   \bar\psi \cdot z = 0,
\end{eqnarray}
which  define  the ${CP}^{(1)}$ model on a sphere $\bar z \cdot z = 1$ 
(that is supersymmetrized with the inclusion of the above fermionic variables $\psi$ and $\bar\psi$).
It is worthwhile to mention that the invariance of the constraints $(\bar z\cdot z - 1 = 0)$ under the
$\mathcal{N} = 2$ SUSY transformations $s_1$ and $s_2$, leads to the constraints $\bar z\cdot \psi = 0$  
and $\bar\psi\cdot z = 0$. Mathematically, these constraints have also been determined by the superspace 
 formalism in [29] which has been concisely mentioned in our Appendix A
for the readers' convenience.

We obtain a bosonic symmetry ($s_\omega$) in the theory which is nothing but 
the anticommutator  $s_\omega = \{s_1, \, s_2\}$ of the $\mathcal{N} = 2$ SUSY 
transformations $s_1$ and $s_2$. The dynamical 
variables of our theory
transform, under $s_\omega$, as follows: 
\begin{eqnarray}
s_\omega \, z =  D_t\,z, \qquad s_\omega \, \bar z = D_t\,\bar z,\qquad
s_\omega \, \psi = D_t\,\psi, \qquad s_\omega \, \bar\psi = D_t\,\bar\psi,
\end{eqnarray}
modulo a factor of $i$. Thus, it is clear that the anticommutator of two SUSY transformations generates 
the ``covariant" time translation for the variables. This is one of the key requirements of a
consistent  $\mathcal{N} = 2$
interacting SUSY theory which is clear from the superspace and the (anti-)chiral 
supervariable formalism (see, Appendix A). Under  the bosonic symmetry transformations (7), the
Lagrangian transforms as: $s_\omega L = \frac{d}{dt} [ L + 2 g a] \equiv 
\frac{d}{dt} [2 (D_t {\bar z}) \cdot (D_t z) + \frac{i}{2}\, \{ \bar\psi \cdot (D_t\psi)
- (D_t {\bar\psi})\cdot \psi \}]$. As a consequence, the action integral 
of our theory remains invariant under $s_\omega$ for the physically well-defined variables.

The existence of continuous symmetries in a theory leads to the derivation of conserved Noether charges 
which turn out to be the  generators for the continuous symmetry transformations.
The continuous and nilpotent  $\mathcal{N} = 2$ SUSY transformations $s_1$ and $s_2$ and their anticommutator 
$s_\omega$ lead to the derivation of the following Noether conserved charges  
\begin{eqnarray}
 Q &=&  \frac{\Pi_z  \cdot \psi}{\sqrt{2}} = \frac{1}{\sqrt{2}}\,\left[2\, D_t\, \bar z 
+ \frac{i}{2}\, (\bar\psi\cdot \psi + 2\, g)\, \bar z + 2\,i\,a\,\bar{z}\,(1 - \bar{z}\cdot z)\right]\cdot \psi \nonumber\\
&\equiv & \frac{1}{\sqrt{2}}\,\left[2\, D_t\, \bar z 
+ \frac{i}{2}\, (\bar\psi\cdot \psi + 2\, g)\, \bar z \right]\cdot \psi,  \nonumber\\
 \bar Q &=&  \frac{\bar\psi \cdot  \,\Pi_{\bar z}}{\sqrt{2}} =\frac{1}{\sqrt{2}}\,  \bar\psi
\cdot \left[2\, D_t \, z - \frac{i}{2}\, (\bar\psi \cdot\psi + 2\, g)\,  z
 - 2\,i\,a\,z\,(1 - \bar{z}\cdot z)\right] \nonumber\\
& \equiv & \frac{1}{\sqrt{2}}\,\bar\psi
\cdot \left[2\, D_t \, z - \frac{i}{2}\, (\bar\psi \cdot\psi + 2\, g)\,  z \right], \nonumber\\
 Q_\omega &=& H  =   2\, (D_t\, \bar{z})\cdot (D_t\,z) 
- \frac{1}{2}\,(\bar{\psi}\cdot \psi + 2g)\,(\bar{\psi}\cdot \psi)\nonumber\\
& \equiv & \frac{1}{2}\,\Pi_z \cdot \Pi_{\bar{z}} - \frac{1}{8}\,\bigl(\bar{\psi} \cdot \psi + 2g \bigr)^2,
\end{eqnarray} 
where $H$ is the Hamiltonian of our present theory. It is evident that,
 we have used the constraint $\bar z \cdot z = 1$ in the shorter version of 
 $Q$ and $\bar Q$ in the above equation (8).

We note  that the  above charges have also been expressed in terms of canonical 
conjugate momenta $\Pi_z$ and $\Pi_{\bar z}$
w.r.t. the dynamical variables   $z$ and $\bar z$. In fact,
it is elementary to check that the canonical momenta (that emerge from Lagrangian (1))
 corresponding to the dynamical variables 
$z, \bar z, \psi$ and $\bar\psi$   are: 
\begin{eqnarray}
 \Pi_z &=& \frac{\partial L}{\partial \,\dot z} = 2\, D_t\, \bar z 
+ \frac{i}{2}\, (\bar\psi\cdot \psi + 2\, g)\, \bar z + 2\,i\,a\,\bar{z}\,(1 - \bar{z}\cdot z), 
\nonumber\\
&\equiv &   2\, D_t\, \bar z 
+ \frac{i}{2}\, (\bar\psi\cdot \psi + 2\, g)\, \bar z, 
\nonumber\\
\Pi_{\bar z} &=& \frac{\partial L}{\partial \,\dot{\bar z}} =  2\, D_t \, z - \frac{i}{2}\, 
(\bar\psi \cdot\psi + 2\, g)\,  z - 2\,i\,a\,z\,(1 - \bar{z}\cdot z), \nonumber\\
& \equiv &  2\, D_t \, z - \frac{i}{2}\, 
(\bar\psi \cdot\psi + 2\, g)\,  z, \nonumber\\
 \Pi_{\psi} & =& \frac{\partial L}{\partial \, \dot{\psi }} \equiv - \frac{i}{2}\, \bar\psi,
\qquad\qquad\qquad\qquad \Pi_{\bar\psi} =  \frac{\partial L} {\partial \,\dot{\bar\psi }} \equiv 
- \frac{i}{2}\, \psi,
\end{eqnarray}
where we have adopted the convention of  left derivative w.r.t. the fermionic superpartners 
$\psi$ and $\bar\psi$ in the computation of $\Pi_\psi$ and $\Pi_{\bar\psi}$. 
The equivalent forms of the canonical momenta ($\Pi_z, \, \Pi_{\bar z}$) have been 
derived by imposing  the constraint $\bar z\cdot z = 1$. We also
note that the Hamiltonian $H$ of our theory [cf. (8)] can also be computed 
by the Legendre transformations\footnote{ We note that the {\it ordering} and the proper {\it signatures}
have been taken into account in our definition of the Legendre  transformation which leads to the derivation 
of the Hamiltonian $H$.} 
$H =  \Pi_z \cdot \dot z + \dot {\bar z} \cdot \Pi_{\bar z} - \Pi_\psi \cdot \dot \psi 
+ \dot {\bar \psi} \cdot \Pi_{\bar\psi} - L$ by exploiting the definition  of canonical conjugate momenta (9)
and the expression for the Lagrangian (1). 
In this derivation, we have to use the constraint  $\bar z\cdot z = 1$ and definitions of 
$D_t \, z, \, D_t \, \bar z$ and $a = -\, 1/2[i\, (\bar  z \cdot \dot z - \dot{\bar z} \cdot z) 
+\, \bar \psi \cdot \psi]$.

The  Noether charges ($Q,\, \bar Q,\, Q_\omega$) are conserved and their conservation law (i.e. $\dot{Q} = \dot{\bar{Q}} = \dot{Q}_\omega = 0$)
can be proven by exploiting the following equations of motion (along with the constraint $\bar z\cdot z = 1$) that
emerge from the Lagrangian (1) of our theory, namely;
\begin{eqnarray}
&& \frac{d\Pi_{\bar{z}}}{dt} - i\,\left[2\,a\, D_t\, z + \frac{\dot{z}}{2}\,(\bar{\psi}\cdot \psi+ 2g) \right] = 0, \nonumber\\ &&
 \frac{d\Pi_z}{dt}\, + i\, \left[2\,a\, D_t\, \bar{z} + \frac{\dot{\bar{z}}}{2}\,(\bar{\psi}\cdot \psi+ 2g) \right] = 0, \nonumber\\
&& D_t\,\psi - \frac{i}{2}\,(\bar{\psi}\cdot \psi+ 2g)\, \psi = 0, \qquad  D_t\, \bar{\psi} + \frac{i}{2}\,(\bar{\psi}\cdot \psi+ 2g)\, \bar\psi= 0,
\end{eqnarray}
where the canonical conjugate momenta ($\Pi_z,\, \Pi_{\bar{z}}$) are 
defined in the equation (9). It is worth pointing out that the above 
charges are conserved on the constrained surface defined by the constraints 
$\bar{z}\cdot z = 1,\, \bar {z} \cdot \psi = 0, \, \bar{\psi}\cdot z = 0$. 
Thus, in the explicit proof of the conservation law, the EOM and constraints
 {\it both} are exploited together (see, Appendix B).

\noindent
\section{Novel discrete symmetry transformations}

It is straightforward to note that the Lagrangian (1) remains invariant 
under the following discrete symmetry transformations, namely;
\begin{eqnarray}
&& z \rightarrow  \mp \,\bar z, \qquad \bar z \rightarrow   \mp\,  z, \qquad
\psi \rightarrow  \mp\, \bar \psi, \qquad \bar\psi \rightarrow  \pm \, \psi, \nonumber\\ && 
t \rightarrow -\, t,\qquad a \rightarrow + a,  \qquad g \to g, 
\end{eqnarray}
where we note that there is a time-reversal (i.e. $t \rightarrow -\,t$) symmetry. 
In other words, we observe that, in reality, the transformation $z \rightarrow  \mp\,\bar{z}$ 
denotes that $z(t) \rightarrow z(-t) = \mp\,{\bar z}^T(t)$ where the superscript
 $T$ on $\bar{z}$ denotes the transpose operation  on $\bar{z}$.
 We have suppressed  the transpose operations  in the transformation  (11). This operation, however, 
 should be taken into account in the rest of the   discrete symmetry
 transformations (e.g. $\psi \rightarrow \mp\,\bar{\psi} \Rightarrow \psi(t)
 \rightarrow \psi(-\,t) = \mp\,{\bar\psi}^T (t),\,   \bar\psi \rightarrow \,{\psi} \Rightarrow \bar\psi(t)
 \rightarrow \bar\psi(-\,t) = \pm {\psi}^T (t),\, a (t) \rightarrow a(-t) =
a (t)$,  etc.).

The above discrete symmetry transformations are  useful and important because they facilitates 
a connection between the {\it two} nilpotent (${s_1}^2 = {s_2}^2 = 0$) 
SUSY transformations $s_1$ and $s_2$ in the following manner, namely; 
\begin{eqnarray}
 s_2  = \pm\; * s_1\; *,
\end{eqnarray} 
where the notation $*$ has been exploited for the discrete 
symmetry transformations (11). The ($\pm$) signs in the above 
equation are dictated by  two successive operations of the 
discrete symmetry transformations, namely;
\begin{eqnarray}
&& *\; (\; *\;   \Phi_1) = \; +\; \Phi_1,  \qquad \qquad \Phi_1 = z, \, \bar z, \nonumber\\
&& *\; (\; *\;   \Phi_2) = \; -\; \Phi_2,  \qquad \qquad \Phi_2 = \; \psi , \; \bar \psi.
\end{eqnarray}
As a side remark, we observe that the bosonic variables $z(t)$ and $\bar z (t)$ have the ($+$)
sign when they are acted upon by two successive discrete symmetry transformations. 
On the contrary, the fermionic variables (i.e. $\psi(t)$ and $\bar \psi (t)$) acquire a ($-$)
sign under the above transformations.
Thus, we have the following relationships:
\begin{eqnarray}
&& s_2\, \Phi_1 = +\; * s_1\; *\; \Phi_1 \;\;\Rightarrow\;\; s_2 = + \;*\; s_1\; *, \qquad\qquad
\Phi_1 = z, \,\bar z,   \nonumber\\
&& s_2 \, \Phi_2 = -\; * s_1\; *\; \Phi_2 \;\;\Rightarrow\;\; s_2 = - \;*\; s_1\; *, 
\qquad\qquad \Phi_2 = \psi, \bar\psi,
\end{eqnarray}
where there is an elegant interplay between the discrete and continuous symmetries of our theory. 
We note that there is existence of a set of reciprocal relationships, namely; 
\begin{eqnarray}
&& s_1 \,\Phi_1 = - \; * s_2\; *\; \Phi_1 \;\;\Rightarrow\;\; s_1 = - \;*\; s_2\; *, \qquad\qquad
\Phi_1 = z, \,\bar z,   \nonumber\\
&& s_1 \, \Phi_2 = + \; * s_2\; *\; \Phi_2 \;\;\Rightarrow\;\; s_1 = + \;*\; s_2\; *, 
\qquad\qquad \Phi_2 = \psi, \bar\psi,
\end{eqnarray} 
which is also true for a duality invariant theory [30].

We end this section with the remarks that the relationships (12) between
 the $\mathcal{N} = 2$ SUSY transformations $s_1$ and $s_2$ are reminiscent
 of the relationships that exist between the exterior and co-exterior derivatives 
of differential geometry (i.e. $\delta = \pm\, * \,d \,*$). 
Thus, it is very interesting to note that the 
relationship (12) provides a physical realization for the above mathematical relationship 
 in the language of the interplay between discrete and 
continuous symmetries of our present theory. We would like to lay emphasis on the fact that we have discussed 
only one set of discrete symmetry transformation in (11). However, there might exist many other useful discrete
symmetry transformations (see, Appendix D) in our theory, too. We shall focus, however,  for our rest of the discussions on (11)
as the discrete symmetry transformations of  our theory. It will be noted that 
the conserved charges ($Q,\, \bar Q,\, Q_\omega$) transform, under the discrete
 symmetry transformations (11), as:
\begin{eqnarray}
* \, Q = -\, \bar Q, \qquad\qquad *\, \bar Q = Q,\qquad \qquad * \, H = H. 
\end{eqnarray}
Thus, we point out  that the transformations of the SUSY charges $Q$ and $\bar Q$
are exactly like the duality transformation of Maxwell's theory of electrodynamics where
$\vec E \rightarrow \vec B, \, \vec B \rightarrow -\, \vec E$ 
(under the Maxwell duality symmetry transformations).
Furthermore, we observe that a couple of successive operations of the discrete symmetry transformation
on the charges yield the following transformations: 
\begin{eqnarray}
*\,(*\, Q) = -\, Q, \qquad \quad *\, (*\, \bar Q) = -\, \bar Q, \qquad \qquad *\,(*\, H) = H.
\end{eqnarray}
The above observation establishes  the fermionic nature of $Q$ and $\bar Q$ as we have 
seen earlier in the case of $\psi$ and $\bar\psi$  
(i.e. $*\,(*\, \psi) = -\, \psi,\; *\,(*\, \bar\psi) = -\, \bar\psi$).
It is worthwhile to point out that the algebra: $Q^2 = {\bar Q}^2 =0,\; \{ Q,\, \bar Q\} = H, \, 
[H, \, Q] = [H, \, \bar Q] = 0$
remains invariant under a couple of  successive discrete symmetry transformations.

\noindent
\section{Operators and their algebra}

We focus on the algebraic structure that is followed by the transformation operators ($s_1,\,s_2,\,s_\omega$)
and their corresponding Noether charges ($Q,\,\bar{Q},\,Q_\omega$). In this respect, as has
already  been discussed, we note that the following is true, namely;
\begin{eqnarray}
&& s_1^2 = 0, \;\qquad s_2^2 = 0,\; \qquad \lbrace s_1, s_2\rbrace = s_\omega \equiv (s_1 + s_2)^2, \nonumber\\
&& \left[s_\omega, s_1\right] = 0,\qquad \qquad [s_\omega, s_2] = 0, \qquad\quad \lbrace s_1, s_2 \rbrace  \neq 0.
\end{eqnarray}
Thus, we observe that the bosonic symmetry operator ($s_\omega$) commutes with all the {\it three} continuous symmetry operators ($s_1,\,s_2,\,s_\omega$) of our interacting $\mathcal{N} = 2$ SUSY QMM. Hence, it behaves 
like the Casimir operator. In exactly similar fashion, it can be seen that
(due to the relationship between the continuous symmetries and generators)
 we have the following: 
\begin{eqnarray}
&& s_1\,Q = \,i\,\lbrace Q,\, Q\rbrace = 0, \;\;\qquad \qquad s_2\, \bar{Q} 
= \,i\,\lbrace \bar{Q},\, \bar{Q}\rbrace = 0,\nonumber\\
&& s_1\,\bar  Q = i\,\lbrace \bar{Q},\, Q \rbrace = i\, H, \qquad\qquad s_2\, Q 
= i\,\lbrace Q, \,\bar{Q}\rbrace = i\, H, \nonumber\\
&& s_\omega\, Q = -\,i\, \left[Q, \,H \right] = 0, \qquad\qquad s_\omega \,\bar{Q} 
= -\,i\, \left[\bar{Q},\, H \right] = 0.
\end{eqnarray}
The above relationships  lead to the realization of one of the
 simplest form of the $\mathcal{N} = 2$ SUSY algebra amongst the generators 
of $(s_1, s_2, s_{\omega})$, namely;
\begin{eqnarray}
&& Q^2 = \frac{1}{2}\, \lbrace Q, \,Q \rbrace = 0, \qquad\; \bar{Q}^2 
= \frac{1}{2}\,\lbrace \bar{Q}, \,\bar{Q} \rbrace = 0,\nonumber\\
&& \lbrace Q,\, \bar{Q} \rbrace = H, \qquad\qquad\quad \left[ H, \,Q \right] 
= \left[ H, \,\bar {Q} \right] = 0,
\end{eqnarray}
where there is {\it no} central extension. Thus, the continuous symmetry 
transformations and their generators  respect the celebrated  $sl(1/1)$ algebra 
(without any central extension). The proof of (19) is algebraically a bit involved. 
Thus, for the readers' convenience, we have explicitly derived these relations in our Appendix C.

Physically, the above algebra can be understood as follows. 
The nilpotency ($Q^2 = 0,$ $ \bar{Q}^2 = 0$) of the charges ($Q,\,\bar{Q}$) 
shows that these charges are fermionic in nature. The vanishing of the 
commutators (i.e. $[Q, \, H] = [\bar Q, \, H] = 0$) demonstrates that the SUSY (i.e. fermionic) 
charges ($Q, \bar Q$) are conserved. Finally, the algebraic structure 
$\{Q, \, \bar Q\} = H$ says that the operation of two consecutive operations  of $s_1$ 
and $s_2$  on any arbitrary variable lead to the ``covariant" time translation [cf. Eqn. (7) of Sec. 2]
of the {\it same} variable. This physical statement is mathematically captured by the
presence of  bosonic symmetry transformation ($s_\omega$) in our theory (i.e. $s_\omega = \{s_1,\, s_2\}$).

The algebraic structures (18) and (20) are reminiscent of the algebra  obeyed  by the
de Rham cohomological operators of differential geometry, namely;
\begin{eqnarray}
d^2 = 0, \quad \delta^2 = 0, \quad \Delta = (d + \delta)^2 = \{d, \, \delta \}, \quad [\Delta, \, d]= 0,\quad
[\Delta, \, \delta] = 0,
\end{eqnarray}
where the operators ($d, \, \delta, \, \Delta$) from a set called the de Rham cohomological operators of differential geometry. 
Thus, we note that the Laplacian operator $\Delta$ behaves like a Casimir operator\footnote{This Laplacian operator is not like the precise
Casimir operator of the Lie algebra (as the Hodge algebra (21) is {\it not} a Lie algebra). Similarly, the algebras
(18) and (20) are {\it also} not Lie algebra.} 
exactly as $s_\omega$ and $H$ behave in our equations (18) and (20). We note that the continuous symmetry 
operators ($s_1, \, s_2, \, s_\omega$) and their generators ($Q, \, \bar Q, \, Q_\omega$)
provide the physical realization of the cohomological operators ($d, \, \delta, \, \Delta$) of differential geometry
in the language of symmetries and conserved charges.

We close this section with the remark that the well-known relation: $\delta = \pm \, *\, d \, *$
(that exists between the co-exterior derivative and exterior derivative) is realized in the language of the
interplay between the discrete and continuous symmetries of our present theory
[cf. (12)]. Thus, as far as the algebraic structure is concerned, we note that there is complete similarity 
amongst the relations (18), (20) and (21) and there is a mapping between 
($d, \delta, \Delta $) and the conserved charges ($Q, \, \bar Q, \, Q_\omega$). 
This identification is, however, still not complete
because there are cohomological properties that are associated with ($d, \delta, \Delta $)
and we have to capture these properties in the language of conserved charges ($Q, \, \bar Q, \, Q_\omega$).
 We shall dwell on these  aspects in our next section.

\noindent
\section{Cohomological connections}

It is a well-known fact that when the de Rham cohomological operators ($d,\, \delta,\,\Delta $) 
act on a given form  (i.e. $f^{(p)}$) of degree 
$p$, the following consequences ensue. First, the degree of this  form increases 
by one when it is acted upon by the exterior derivative $d$ (i.e. $d \,f^{(p)}\sim f^{(p + 1)} $). 
On the contrary, the operation of $\delta$ on the same form (i.e. $f^{(p)}$) 
lowers the degree of the form by one (i.e. $\delta \,f^{(p)}\sim f^{(p - 1)} $ ). 
It turns out that the operation of the Laplacian operator does {\it not } 
change the degree of the form at all (i.e. $\Delta \,f^{(p)} \sim f^{(p)}$). These properties  are the essential ingredients when 
we discuss the cohomological aspects of a differential form w.r.t. the de Rham 
cohomological operators. We have to find the analogues of these observations 
in the language of the symmetry properties and conserved charges of the interacting 
$\mathcal{N} = 2$ SUSY QMM of our present endeavor and establish a proper relationship.

We have noted that $H$ is like the Casimir operator for the $sl(1/1)$
algebra: $Q^2 = {\bar{Q}}^2 = 0$,  $\lbrace Q, \bar{Q}\rbrace = H,\,
\left[H, Q \right] = \left[H, \bar{Q}\right] = 0$. As a consequence
of the latter two relations, it is evident that $H^{-1}\,Q = Q\,H^{-1}$
and $H^{-1}\,\bar{Q} = \bar{Q}\,H^{-1}$ (where we assume that $H^{-1}$ 
is properly defined for the Hamiltonian of our theory). We note that the 
following relations turn out to be true due to the validity of the $sl(1/1)$ algebra,  namely;
\begin{eqnarray}
&& \left[ \frac{Q\,\,\bar{Q}}{H},\, Q \right] = + \, Q, \qquad\qquad \left[\frac{Q\,
\bar{Q}}{H},\, \bar{Q} \right]
 = - \, \bar{Q}, \nonumber\\
&& \left[\frac{\bar{Q}\,Q}{H}, \,Q \right] = -\, Q, \qquad
\qquad \left[\frac{\bar{Q}\,Q}{H},\,\bar{Q} \right] = + \, \bar{Q},
\end{eqnarray}
where it has been taken for granted that $H^{-1}\, Q = Q\, H^{-1}$ 
and $H^{-1}\, \bar{Q} = \bar{Q}\, H^{-1}$ are very much true. Let us  define now a state 
$\mid \chi>_p $ (in the total quantum Hilbert space of states), with an eigenvalue equation
w.r.t. the Hermitian operator (${Q\,\bar{Q}}/{H}$),  as:
\begin{eqnarray}
\left(\frac{Q\,\bar{Q}}{H}\right)\,\mid \chi>_p \,= p\,\mid \chi >_p,
\end{eqnarray}
where $p$ is the eigenvalue which is {\it real} because $({Q\,\bar{Q}}/{H})$
operator is Hermitian. Now, we can discuss about the impact of the algebra (22) on the state 
$\mid \chi>_P$ which could be useful in the description of the cohomological aspects of 
states. In this context, we shall utilize the beauty of equation (22) as it
behaves like the ladder operators of quantum mechanics. For instance, using 
the top two relationships from (22), we find  that
\begin{eqnarray}
&& \left(\frac{Q\,\bar{Q}}{H}\right)\, Q \mid \chi >_p\, = (p + 1)\,Q\mid \chi >_p, 
\nonumber\\
&& \left(\frac{Q\,\bar{Q}}{H}\right)\, \bar{Q} \mid \chi >_p \,= (p - 1)\,\bar{Q} 
\mid \chi >_p, \nonumber\\
&& \left(\frac{Q\,\bar{Q}}{H}\right)\, H \mid \chi >_p \,= p \,H \mid \chi >_p. 
\end{eqnarray}
Thus, it is crystal clear that the states $Q\, |\chi >_p, \, \bar{Q} \,|\chi>_p$ 
and $H \,|\chi>_p$ have the eigenvalues $(p + 1), (p - 1)$ and $p$, respectively, 
w.r.t. the Hermitian operator $\left({Q\,\bar{Q}}/{H}\right)$.
These consequences are {\it exactly} same as the results that emerge from the 
operations of the cohomological operators $(d, \delta, \Delta)$ 
on a form $f^{(p)}$ of degree $p$ (because we do obtain  the forms of degree 
$(p + 1), (p - 1)$ and $p$, respectively) due to their operations.

Now we are in a position to exploit the mathematical beauty and power of the lower 
{\it two} entries of equation (22). If we define an eigenstate $\mid \xi>_q$ (in the 
total quantum Hilbert space of states) w.r.t. the Hermitian operator
$\left({\bar{Q}\,Q}/{H}\right)$, as 
\begin{eqnarray}
\left(\frac{\bar{Q}\,Q}{H}\right)\,\mid \xi>_q \,=  q\mid \xi>_q,
\end{eqnarray} 
it is evident that the following consequences  ensue,  namely;
\begin{eqnarray}
&& \left(\frac{\bar{Q}\,Q}{H}\right)\, \bar{Q} \mid \xi>_q \,= (q + 1)\,\bar{Q}\mid \xi>_q, \nonumber\\
&& \left(\frac{\bar{Q}\,Q}{H}\right)\, Q \mid \xi>_q\, = (q - 1)\,Q \mid \xi>_q, \nonumber\\
&& \left(\frac{\bar{Q}\,Q}{H}\right)\, H \mid \xi>_q \,= q \,H \mid \xi>_q,
\end{eqnarray}
which shows that the eigenvalues of states  $\bar{Q}\mid \chi>_q,\, Q\mid \chi>_q $ and 
$H\mid \chi>_q $ are $(q + 1),(q - 1)$ and $q$, respectively, w.r.t. the Hermitian operator 
$\left({\bar{Q}\,Q}/{H}\right)$. Once again, we note that these consequences are exactly 
same as the operations of the cohomological operators ($ d,\delta,\Delta $) on the differential
form of degree $q$. Hence, these mathematical operators can be realized in the language of 
conserved charges of our present theory and explained below.

Due to the beauty  of equations (24) and (26), we observe 
 that if an eigenstate $\mid \chi>_p$ has an eigenvalue  $p$ w.r.t. the
Hermitian operator ($Q\,\bar{Q}/{H}$), the states $Q | \chi>_p, \bar{Q} | \chi>_p$
and $H\mid \chi>_p$ would have the eigenvalues  $(p + 1),(p - 1)$ and $p$, respectively. Now, if we identify
a corresponding differential form $f^{(p)}$ of degree $p$, we know that $d\,f^{(p)}, \delta\,f^{(p)}$ and 
$\Delta\,f^{(p)}$ would have the degrees $(p + 1),(p - 1)$ and $p$, respectively. Thus, we conclude that 
we have a mapping:
\begin{equation}
(Q, \bar{Q}, H)\; \Longleftrightarrow \;(d,\delta,\Delta),
\end{equation}
between the two kinds of operators. On the other hand, if we identify the eigenstate $ \mid\xi>_q$ 
with eigenvalue $q$, w.r.t. the Hermitian operator (${\bar{Q}\,Q}/H$), we observe that the quantum states 
$\bar{Q}\mid \xi>_q,\, Q\mid \xi>_q $ and $H\mid \chi>_q $ would have their eigenvalues 
$(q + 1),(q - 1)$ and $q$, respectively. Similarly, the forms $d\,f^{(q)}, \delta\,f^{(q)}$  
and $\Delta\,f^{(q)}$ would 
carry the degrees $(q + 1),(q - 1)$ and $q$, respectively, 
if we start with a form $f^{(q)}$ of degree $q$.  Thus, we have the mapping of the operators:
\begin{equation}
(\bar{Q}, Q, H) \Longleftrightarrow (d,\delta,\Delta),
\end{equation}
which are defined in two different spaces. Whereas, the operators ($\bar{Q}, Q, H$) are defined in the 
quantum Hilbert space of states, the operators $(d, \delta,\Delta)$ are defined in the space of forms
on a given manifold. All the above discussions imply that there are {\it two} realizations of
($d, \, \delta, \, \Delta$) in the language of conserved charges of our present $\mathcal{N} = 2$
SUSY theory. Depending on the starting quantum state w.r.t. a given Hermitian operator, we
have the mappings: $(Q, \bar{Q},  H) \Longleftrightarrow (d,\delta,\Delta)$ and/or 
$(\bar{Q}, Q, H) \Longleftrightarrow (d,\delta,\Delta)$.

\section{Nilpotent $\mathcal{N} = 2$ SUSY transformations: Supervariable approach
and geometrical interpretations }

\noindent
First of all, we focus on the derivation of $s_1$ within the framework of the supervariable 
approach. In this regards, we generalize the ordinary  variables $\left( z(t), \bar{z}(t), \psi(t), 
\bar{\psi}(t) \right)$ to their counterparts supervariables on a (1, 1)-dimensional chiral 
super-submanifold  parametrized by ($t, \theta$).
This super submanifold is a part of the general (1, 2)-dimensional 
supermanifold on which our theory is generalized. 
Thus, chiral supervariables and their expansions are:
\begin{eqnarray}
&& z(t) \longrightarrow Z(t, \theta) = z (t) + \theta\, f_1(t),\nonumber\\
&& \bar{z}(t) \longrightarrow \bar{Z}(t, \theta) = \bar{z}(t) + \theta\, f_2(t),\nonumber\\
&& \psi(t) \longrightarrow \Psi(t, \theta) = \psi(t) + i\,\theta\,b_1(t), \nonumber\\
&& \bar{\psi}(t) \longrightarrow \bar{\Psi}(t, \theta) = \bar{\psi}(t) + i\, \theta\, b_2(t),
\end{eqnarray}
where secondary variables ($f_1(t), f_2(t)$) and $\left( b_1 (t), b_2 (t) \right)$ are fermionic 
and bosonic in nature as is evident from the fermionic nature (i.e. $Q^2 = 0$) of Grassmannian variable $\theta$.

To determine these secondary variables in terms of the basic variables, we have to invoke  SUSY
invariant restrictions (SUSYIRs). In this connection, we note that the following useful quantities remain 
invariant under $s_1$, namely;
\begin{equation}
s_1\,(\psi) = 0,\quad s_1\,(\bar{z})  = 0, \quad s_1\, (z^T \cdot \psi)  = 0, 
\quad s_1\, \left[2\,D_t\,{\bar{z}}\cdot z  + i\,\bar{\psi}\cdot \psi\right] = 0,
\end{equation}
where $z^T(t) \cdot \psi(t) = z_1\,\psi_1 + z_2\,\psi_2$. 
The SUSYIRs demand that all the quantities that are present in the above  brackets should remain 
independent of the ``soul" coordinate\footnote{In the old literature (see, e.g. [27]),
the Grassmannian variables have been christened as the ``soul" coordinates and the ordinary
spacetime variables have been called as the ``body" coordinates.} $\theta$ when they are generalized onto the 
(1, 1)-dimensional chiral super submanifold.  In other words, we have the following 
equalities as the SUSYIRs, namely;
\begin{eqnarray}
&& \Psi (t, \theta) = \psi (t) \;\Longrightarrow \; b_1 (t) = 0,\qquad \qquad
 \bar{Z}(t, \theta) = \bar{z}(t) \; \;\Longrightarrow \; f_2 (t)  = 0,\nonumber\\
&& Z^T (t, \theta) \cdot \Psi (t, \theta) = z^T (t) \cdot \psi (t), \nonumber\\
&& 2\,D_t \bar{Z}(t, \theta)\cdot Z (t, \theta) + i\,\bar{\Psi} (t, \theta) \cdot \Psi (t, \theta) = 
2\, D_t\,\bar{z}(t) \cdot z(t) + i\, \bar{\psi}(t) \cdot \psi (t). 
\end{eqnarray}
From the relationship number three,  we obtain the relationship that $f_1 (t) \propto \psi (t) $.
In this regards, we make a choice and take $f_1 (t) =  {\psi (t)}/{\sqrt{2}}$. This choice, entails upon 
$b_2 (t) = {2\,D_t\,\bar z (t)}/ {\sqrt {2}}$. Plugging in these values into the expansions, 
we obtain the following
\begin{eqnarray}
&& Z^{(1)}(t, \theta) = z (t) + \theta\, \left(\frac{\psi (t)}{\sqrt{2}} \right) \equiv z (t) 
+ \theta\, \left(s_1\,z (t) \right), \nonumber\\
&& \bar{Z}^{(1)}(t, \theta) = \bar{z} (t) + \theta\, (0)  \equiv \bar{z}(t) 
+ \theta\, \left(s_1\,\bar{z} (t) \right), \nonumber\\
&& \Psi^{(1)}(t, \theta) = \psi(t) + \theta\,(0)  \equiv \psi (t)
 + \theta\, \left(s_1\,\psi (t) \right), \nonumber\\
&& \bar{\Psi}^{(1)}(t, \theta) = \bar{\psi} (t)
 + \theta\, \left( \frac{2\,i\,D_t\,\bar{z}(t)}{\sqrt{2}} \right) \equiv \bar{\psi} (t)
 + \theta\, \left(s_1\,\bar{\psi} (t) \right),
\end{eqnarray} 
where the superscript $(1)$ on the supervariables stand for the chiral supervariables that have been 
obtained after the application of SUSYIRs in (31). A close look at (32) shows that we have already
obtained the SUSY transformations $s_1$.

The above chiral expansions of the supervariables provide the geometrical meaning 
to the SUSY transformations $s_1$ because we note that
\begin{eqnarray}
\frac{\partial}{\partial\theta}\,\Omega^{(1)} (t, \theta) = s_1\, \omega (t) 
\equiv \pm\,i\,\left[\omega (t), Q \right]_{\pm},
\end{eqnarray}
where $\Omega^{(1)} (t, \theta)$ is the generic chiral  supervariable that has been obtained
after the application of SUSYIRs in (31) and $ \omega (t)$ is the generic ordinary 
variable. The subscripts $(\pm)$, on the square bracket, denote the existence of  (anti)commutator for the 
given variable $ \omega (t)$ being (fermionic)bosonic  in nature. Thus, we observe that the 
operators $(\partial_\theta, s_1, Q)$ are inter-connected. Geometrically, the SUSY 
transformation $s_1$ on an ordinary variable is equivalent to the translation of the 
corresponding supervariable (obtained after SUSYIRs) along the Grassmannian $\theta$-direction
of the (1, 1)-dimensional chiral super-submanifold. Furthermore, two successive translations
along this Grassmannian direction captures the nilpotency $( s_1^2 = 0, Q^2 = 0)$
of $s_1$ and $Q$   which is  nothing other than the  nilpotency ($\partial^2_{\theta} = 0$) property of 
the translational generator $\partial_{\theta}$
 along the $\theta$-direction.

To derive the SUSY transformations $s_2$, we generalize the ordinary variables $(z(t), 
\bar{z}(t)$, $\psi(t), \bar{\psi}(t))$ onto (1, 1)-dimensional anti-chiral super-submanifold
that is characterized by $(t, \bar\theta)$. In other words, we have the following:
\begin{eqnarray}
&& z(t) \; \longrightarrow \;  Z(t, \bar\theta) = z(t) +  \bar\theta\, f_3(t),\nonumber\\
&& \bar z(t) \; \longrightarrow \; \bar Z(t, \bar\theta) = \bar z(t) +  \bar\theta\, f_4(t),\nonumber\\
&& \psi(t) \;\longrightarrow \; \Psi (t, \bar\theta) = \psi (t)  + i\,\bar \theta\, b_3 (t), \nonumber\\
&& \bar\psi (t)\; \longrightarrow  \; \bar\Psi (t, \bar\theta) = \bar\psi (t)  + i\, \bar\theta\, b_4 (t),
\end{eqnarray} 
where ($f_3 (t), f_4 (t) $) and ($b_3 (t), b_4 (t)$) are the pair of fermionic  and bosonic 
secondary variables that have to be determined by exploiting the SUSYIRs on the anti-chiral
supervariables. In this connection,
it can be checked that the following is true, namely;
\begin{equation}
s_2\,(\bar\psi) = 0,\quad s_2\,({z})  = 0, \quad s_2\, (\bar z \cdot {\bar\psi}^T)  = 0, 
\quad s_2\, \left[2\,{\bar{z}}\cdot D_t\, z  - i\,\bar{\psi}\cdot \psi\right] = 0.
\end{equation}
Thus, the above invariant quantities would remain independent of the 
``soul" coordinate $\bar \theta$ when they are
generalized onto the anti-chiral super sub-manifold. 
In other words, we have the following SUSYIRs in the form 
of the equalities, namely;
\begin{eqnarray}
&& Z (t, \bar\theta) = z (t), \qquad \bar\Psi (t, \bar \theta)
= \bar\psi(t), \qquad  \bar Z (t, \bar\theta)\cdot {\bar\Psi}^T (t, \bar\theta) 
= \bar z (t)\cdot {\bar\psi}^T (t), \nonumber\\
&& 2\,{\bar Z} (t, \bar\theta) \cdot D_t\,Z (t, \bar\theta) - i \, {\bar\Psi} (t, \bar\theta) 
\cdot \Psi (t, \bar\theta)  =  2\,{\bar z}(t)\cdot  D_t\,z(t) - i\,{\bar\psi}(t)\cdot\psi(t),
\end{eqnarray}
which lead to the  determination of the secondary variables as 
\begin{eqnarray}
\quad f_3(t) = 0, \qquad b_4(t) = 0, \qquad f_4 (t)
 = \frac{\bar\psi (t)}{\sqrt 2}, \qquad b_3 (t) = \frac{2\, D_t\,z (t)}{\sqrt 2}.
\end{eqnarray}
The substitution of these secondary variables into the expansions (34) lead to the following
\begin{eqnarray}
&& Z^{(2)}(t, \bar\theta) = z(t) + \bar \theta\,(0) \equiv z(t) 
+ \bar\theta \,(s_2\, z),\nonumber\\
&& {\bar Z}^{(2)}(t, \bar\theta) = \bar z(t) +  \bar\theta\, \left(\frac{\bar\psi}{\sqrt 2} \right) \equiv \bar z(t) 
+ \bar\theta \,(s_2\, \bar z),\nonumber\\
&& \Psi^{(2)} (t, \bar\theta) = \psi (t)  + \bar\theta\,\left(\frac{2\,i\,D_t\,{ z}}{\sqrt 2} \right)
 \equiv \psi(t) + \bar\theta\, (s_2\, \psi), \nonumber\\
&& \bar\Psi^{(2)} (t, \bar\theta) = \bar\psi (t)  + \, \bar\theta\,(0)  
\equiv \bar\psi (t)  + \, \bar\theta\, (s_2\, \bar\psi),
\end{eqnarray} 
where the superscript $(2)$ denotes the expansions of the supervariables after the application of SUSYIRs
in (36). The geometrical meaning of the transformations $s_2$ and its nilpotency can be given in terms of the 
translational generator $\partial_{\bar\theta}$ along $\bar\theta$-direction of the (1, 1)-dimensional
anti-chiral super submanifold along exactly  similar lines  as the ones that have been given for $s_1$ 
after equation (32) and (33).

The conserved charges $Q$ and $\bar Q$  can also be generalized  onto the (anti-)chiral super sub-manifolds
where the supervariables would be taken 
after the application of SUSYIRs. These expressions in terms of the $\partial_\theta, \partial_{\bar \theta}, d \theta,
d \bar\theta$ and (anti-)chiral supervariables are
\begin{eqnarray}
 Q &=&\frac{\partial}{\partial \theta}\, \Big[2\,D_t {\bar Z}^{(1)}(t, \theta)
\cdot Z^{(1)}(t, \theta)\Big]  
 \equiv  \frac{\partial}{\partial \theta}\, \Big[2\,D_t {\bar z}(t)
\cdot Z^{(1)}(t, \theta)\Big], \nonumber\\ 
 &=& \int d\theta\, \Big[2\, D_t{\bar Z}^{(1)}(t, \theta)
\cdot Z^{(1)}(t, \theta)\Big] \equiv \int d\theta\, \Big[2\,D_t {\bar z}(t)
\cdot Z^{(1)}(t, \theta)\Big],  \nonumber\\
Q &=& \frac{\partial}{\partial \theta}\, \Big[- i\,{\bar \Psi}^{(1)}(t, \theta)
\cdot \Psi^{(1)}(t, \theta)\Big]  
 \equiv \frac{\partial}{\partial \theta}\, \Big[- i\,{\bar \Psi}^{(1)}(t, \theta)
\cdot \psi(t)\Big], \nonumber\\ 
 &=& \int d\theta\, \Big[- i\,{\bar \Psi}^{(1)}(t, \theta)
\cdot \Psi^{(1)}(t, \theta)\Big]  
\equiv  \int d\theta\, \Big[- i\,{\bar \Psi}^{(1)}(t, \theta)
\cdot \psi(t)\Big],  \nonumber\\
\bar Q &=&\frac{\partial}{\partial \bar\theta}\, \Big[2\,{\bar Z}^{(2)}(t, \bar\theta)
\cdot D_t { Z}^{(2)}(t, \bar\theta)\Big] 
 \equiv  \frac{\partial}{\partial \bar\theta}\, \Big[2\,{\bar Z}^{(2)}(t, \bar\theta)
\cdot D_t { z}(t)\Big], \nonumber\\ 
&=& \int d \bar\theta\, \Big[2\,{\bar Z}^{(2)}(t, \bar\theta)
\cdot D_t { Z}^{(2)}(t, \bar\theta)\Big] 
 \equiv   \int d \bar\theta\, \Big[2\,{\bar Z}^{(2)}(t, \bar\theta)
\cdot D_t { z}(t)\Big],  \nonumber\\
 \bar Q &=&\frac{\partial}{\partial \bar\theta}\, \Big[+ i\,{\bar \Psi}^{(2)}(t, \bar\theta)
\cdot \Psi^{(2)}(t, \bar\theta)\Big]  
 \equiv  \frac{\partial}{\partial \bar\theta}\, \Big[+ i\,{\bar \psi}(t)
\cdot \Psi^{(2)}(t, \bar\theta)\Big], \nonumber\\ 
&=& \int d \bar\theta\, \Big[+ i\,{\bar \Psi}^{(2)}(t, \bar\theta)
\cdot \Psi^{(2)}(t, \bar\theta)\Big] 
 \equiv  \int d \bar\theta\, \Big[+ i\,{\bar \psi}(t)
\cdot \Psi^{(2)}(t, \bar\theta)\Big].
\end{eqnarray}
The nilpotency of $\partial_\theta$ and $\partial_{\bar\theta}$
(i.e. $\partial_\theta^2 = 0, \,\partial_{\bar\theta}^2 = 0$) implies that the nilpotency of charges
$Q$ and $\bar Q$ (\i.e. $\partial_\theta\, Q = 0, \partial_{\bar\theta}\, \bar Q = 0$). 
Furthermore, when we express the above expressions for the charges 
 $Q$ and $\bar Q$  in terms of the  symmetry transformations $s_1$ and $s_2$ and dynamical 
variables, we observe that 
\begin{eqnarray}
 && Q = s_1 \Big(2\, D_t {\bar z}\cdot z\Big) \equiv s_1\, \Big( - i\, \bar\psi \cdot\psi\Big), \qquad
 \bar Q = s_2 \Big(2\, {\bar z}\cdot D_t z\Big) \equiv s_2\, \Big( + i\, \bar\psi \cdot\psi\Big).
\end{eqnarray}
The nilpotency of charges $Q$ and $\bar Q$ can be readily proven by using the constraints $\bar\psi \cdot z = 0$ 
and $\bar z \cdot \psi = 0$ that have been taken into considerations in our present endeavor.

We can also capture the invariance of the Lagrangian (1) in terms of the (anti-) chiral supervariables that are  
obtained after  the impositions of the SUSYIRs as given below
\begin{eqnarray}
 L \;\rightarrow \; {\tilde L}^{(ac)} & = & 2\, D_t {\bar Z}^{(2)} \cdot D_t { Z}^{(2)} 
  +  \frac{i}{2}\, \left[\bar\Psi^{(2)} \cdot D_t {\Psi}^{(2)}  
- D_t {\bar\Psi}^{(2)} \cdot {\Psi}^{(2)}\right] - 2\,g\, a ,\nonumber\\
 L \; \rightarrow \; {\tilde L}^{(c)} & = & 2\,D_t {\bar Z}^{(1)} \cdot D_t { Z}^{(1)}  
 +  \frac{i}{2}\, \left[\bar\Psi^{(1)} \cdot D_ t {\Psi}^{(1)}
 -  D_t {\bar\Psi}^{(1)} \cdot {\Psi}^{(1)}\right]- 2 \, g \, a, 
\end{eqnarray}
where the superscripts $(c)$ and $(ac)$ denote the chiral and anti-chiral nature of the Lagrangians 
${\tilde L}^{(c)}_0$  and ${\tilde L}^{(ac)}_0$, respectively.
The superscripts $(1)$ and $(2)$ on the supervariables correspond  to the expansions (32) and (38).
  Mathematically,  we  observe  that
\begin{eqnarray}
\frac{\partial}{\partial\theta}\; \Big[{\tilde L}^{(c)} \Big] =  \frac{d}{dt}\Big(\frac{D_t {\bar z} 
\cdot \psi}{\sqrt 2} \Big) \, \equiv\, s_1\, L, \nonumber\\
\frac{\partial}{\partial\bar\theta}\; \Big[{\tilde L}^{(ac)} \Big] =  
\frac{d}{dt}\Big(\frac{\bar\psi\cdot {D_t z}}{\sqrt 2}  \Big) \,\equiv \,s_2\, L.
\end{eqnarray}
The relation (42) provides the geometrical interpretation for the SUSY invariances
of the Lagrangian (1). This can be stated  that  in the following manner.
The (anti-)chiral super Lagrangians (41) are the sum of composite supervariables that have been obtained
after the application of the SUSYIRs. The translations of these super Lagrangians along the $\bar\theta$
and $\theta$-directions of the (anti-)chiral super-submanifolds are such that they produce the 
ordinary total time derivatives [cf. (42)] in the ordinary space thereby
 leaving the action integral ($S = \int dt\, L$) invariant as the 
physical variables vanish off at $t = \pm \infty $ on physical grounds.

\section{Conclusions}

The central theme of our present investigation has been to establish that the {\it interacting}
$\mathcal{N} = 2$ SUSY QMM of the motion of an electron on a sphere, in the background of a 
magnetic monopole, provides a tractable SUSY model for the Hodge theory. We have accomplished 
this goal in our present endeavor because we have shown that the discrete and continuous 
symmetries of our present theory are such that they {\it together} provide the physical 
realizations of the de Rham cohomological operators of differential geometry. A few of 
the specific properties of these cohomological operators are {\it also} captured in the 
language of the conserved charges of our present theory in the quantum Hilbert space.

Some of the subtle issues of our present model are the observations that the constraints
(e.g. $\bar{z}\cdot z = 1, \bar{\psi} \cdot z = 0, \bar{z} \cdot \psi = 0$) and the 
equations of motion (10) {\it together} play very important roles in the proof of the conservation 
laws as well as  in the determination of the $sl(1/1)$ algebra from the symmetry principles 
and conserved charges. In particular, as discussed in Appendix C, the proof of $\lbrace Q,\, \bar{Q}
\rbrace = H$, from the symmetry transformations $s_1\,\bar{Q}$ and $s_2\,Q$, requires the
mathematical beauty and power of the constraints as well as the equations of motion. It is 
gratifying to observe that the {\it interacting} version of our earlier work on the {\it free}    
$\mathcal{N} = 2$ SUSY QMM [5] also turns out to be the model for the Hodge theory
as the symmetries of the theory provide the physical realizations of ($d, \delta, \Delta$).

We have derived the $\mathcal{N} = 2$ SUSY transformations $s_1$ and $s_2$
by exploiting the supervariable approach where we have been theoretically 
compelled to consider {\it only} the (anti-)chiral supervariables. This is due to 
the fact that $\mathcal{N} = 2$ SUSY transformations are nilpotent ($s^2_1 = s_2^2 = 0$) but they are 
{\it not} absolutely anticommuting (i.e. $s_1\, s_2 + s_2\, s_1 \ne 0$). Thus, eventhough 
our theory is generalized onto a specific (1, 2)-dimensional supermanifold, we have {\it not}
taken  the {\it full} expansions of the supervariables along 
($1,\, \theta,\,\bar\theta,\,\theta\, \bar\theta$)-directions of the supermanifold (because the full
expansions would automatically imply the  validity of absolute anticommutativity).
We have also provided the geometrical basis for nilpotency and invariance of the Lagrangian within the 
framework of our supervariable approach [5,18-21].

Besides our present work on the $\mathcal{N} = 2$ SUSY case of the charge-monopole system, there are many 
interesting works (see, e.g. [31-40]) on this particular system as well as that of  the charge-dyon system
(and their very beautiful and diverse   super-extensions). For instance, 
it has been shown, in a couple of very interesting papers [31,32], that our
present system exhibits a hidden conical dynamics. The elaborate discussions on the integrability properties
of this system, its systematic  SUSY generalizations, its connection with the reflectionless 
potentials, etc., have also attracted a great deal of interest amongst theoretical
physicists of  very different backgrounds (see, e.g. [31-40] for more details). We would like to lay emphasis on the fact that, in our present investigation, we have concentrated {\it only} on the algebraic structures of the symmetries and corresponding conserved charges for this system. However, a whole range of other directions remain to be explored and these remain precisely 
an open set of interesting problems  for the future investigations [41].

Our present $\mathcal{N} = 2$ SUSY QMM is {\it special} in the   sense  that there are no singularities 
in our theory mainly because the constraints $\bar{z} \cdot z = 1,\, \bar{\psi} \cdot z = 0,
\,\bar{z}\cdot \psi = 0$ (and their time derivatives) {\it do} play important role in avoiding 
them. The other good feature of our present model is the observation that it can be generalized 
to its counterpart $\mathcal{N} = 4$ version [28]. Thus, it would be a very nice idea to look 
for the physical realizations of the cohomological operators in the case of $\mathcal{N} = 4$ and 
$\mathcal{N} = 8$ versions of our present model. We speculate that this understanding might be 
useful for us in the study of  $\mathcal{N} = 2,4,8$ SUSY gauge theories within the framework of BRST 
formalism where we can also look for their topological nature.
Such speculations  are based on our experience in this kind of study in the context
of 2D (non-)Abelian gauge theories [11] which have been shown to present a {\it new}
class of topological field theories. These new topological field theories
have been shown to possess the Lagrangian density that look like the Witten-type topological
field theories. However, their (anti-)BRST and (anti-)co-BRST symmetries are just like the Schwarz-type
topological field theories as they do not incorporate the shift symmetries
which are the hallmark of a {\it typical} Witten-type  topological field theories.   
We are presently busy with these theoretical  ideas and we 
shall report about our progress in our future publications. \\

\noindent
{\bf Acknowledgements:} 
Two of us (SK and DS) would like to gratefully acknowledge
financial support from UGC, Government
of India, New Delhi, under their SRF-schemes.\\

\vskip .5cm

\noindent
{\bf{\large{~~~~~~~~~~~~~~~~Appendix A: On Superspace Formalism}}}\\

\vskip .5cm

\noindent
We invoke here some of the essential ingredients of the superspace formalism to 
clarify some of the expressions and equations   that have been used in our main body of the text. 
In this connection, we define the supercovariant derivatives ($D, \, \bar D$) for the {\it interacting}  
$\mathcal{N} = 2$  SUSY quantum mechanical model as follows (see, e.g. [29])
\[D = \partial_\theta - i\, \bar\theta\, D_t, \qquad\qquad\qquad \bar D 
= \partial_{\bar\theta} - i\, \theta\, D_t. \eqno (A.1)\]
The above operators lead to the definition as well as expansions of the  chiral ($\Phi (t, \theta, \bar\theta)$)  
and anti-chiral ($\bar \Phi (t, \theta, \bar\theta)$) supervariables  as:
\[ \Phi (t, \theta, \bar\theta)   =  z (t )+ \theta \, \psi(t) - i \,\theta\, \bar\theta\, D_t \, z(t), 
\quad\quad(D_t \, z = \dot z  - i\, a\, z), \]
\[\bar\Phi (t, \theta, \bar\theta) = \bar z(t) - \bar\theta \,\bar \psi(t) + i \,\theta\, \bar\theta\, 
D_t \,\bar z(t), \quad\quad (D_t \, \bar z = \dot {\bar z}  + i\, a\, \bar z), \eqno (A.2)\]
which satisfy  ${\bar D}\,\Phi (t, \theta, \bar\theta) = 0$ and ${ D}\, \bar\Phi (t, \theta, \bar\theta) = 0$.
In the $CP^{(1)}$ model approach, we have the following supersymmetric constraint
in terms of $\Phi(t, \theta, \bar\theta)$ and  $\bar\Phi(t, \theta, \bar\theta)$, namely;
\[\bar\Phi(t, \theta, \bar\theta)\cdot\Phi(t, \theta, \bar\theta) - 1 = 0, \eqno (A.3) \]
for the description of the motion of an electron  on a sphere ($\bar z \cdot z = 1$) 
under background of a monopole. Setting the coefficients of $\theta,\, \bar\theta, \, \theta\bar\theta$ 
and constant term equal to zero in (A.3), we obtain the following expressions for the constraints and $a$, namely;
\[\bar z \cdot z = 1, \quad \bar z \cdot \psi = 0, \quad   \bar\psi \cdot z = 0,
\quad a = - \, \frac{[i\, (\bar z \cdot \dot z - \dot{\bar z} \cdot z) 
+  (\bar\psi \cdot\psi)]}{2\, (\bar z \cdot z)}, \eqno (A.4) \]
which have been taken into account in our present endeavor. We have $\mathcal{N} = 2$ SUSY generators for our  
$\mathcal{N} = 2$ {\it interacting} quantum mechanical model as: 
\[ Q = \frac{1}{\sqrt 2} \left(\partial_\theta + i\, \bar\theta\, D_t \right),
\quad\qquad\qquad \bar Q = \frac{1}{\sqrt 2} \left(\partial_{\bar\theta} 
+ i\, \theta\, D_t \right). \eqno (A.5) \]
Together, the above operators satisfy  one of the simplest $\mathcal{N} = 2$ SUSY algebra, as
\[\{Q,  Q\} \equiv Q^2 = 0,  \qquad \{\bar Q,  \bar Q\} 
\equiv {\bar Q}^2 = 0,\qquad \{Q, \bar Q\} = \, i \, D_t.  \eqno (A.6) \]
The last entry shows that two successive $\mathcal{N} = 2$ SUSY transformations 
(corresponding to $s_1$ and $s_2$) on a variable   leads to the covariant ``time" translation
on the same variable  
which has been demonstrated to be true in our Sec. 2. In fact, from equation
(7), it is clear that the bosonic symmetry ($s_{\omega} = \{s_1\, s_2\}$) is such that it transforms
the dynamical variable of our present theory to their ``covariant'' time derivative 
modulo a factor of $i$.\\

\vskip .5cm

\noindent 
{\bf{\large{~~~~~~~~~Appendix B: Conservation law for the charges}}}

\vskip .5cm
\noindent
In this Appendix, we perform some explicit computations that are connected 
with the proof of the conservations of charges $Q$ and $\bar{Q}$. 
In this connection, first of all, we take the straightforward time derivative on the charge $Q$ as follows 
\[\frac{dQ}{dt} = \frac{1}{\sqrt{2}}\,\left[\frac{d \Pi_z}{dt}\cdot \psi 
+ {\Pi_z} \cdot \frac{d \psi}{dt}  \right]. \eqno (B.1) \]
Using the equations of motion from (10), we obtain the  following expressions  for $\dot{Q}$:
\[ \dot{Q} = \frac{1}{\sqrt{2}} \Big[ -\,i\,\left \{2\,a\, D_t\, \bar{z}  + \frac{1}{2}\,(\bar{\psi} \cdot \psi 
+ 2g)\,\dot{\bar{z}} \right \} \cdot \psi + \left \{2 \,D_t\,\bar{z} + 
\frac{i}{2}\,(\bar{\psi} \cdot \psi + 2g)\,\bar{z} \right \} \cdot \dot\psi \Big ]. \eqno (B.2) \]
At this stage, we use the equation of motion w.r.t. $\bar{\psi}$ variable which leads to 
\[ \dot{\psi} = \frac{i}{2}\,(\bar{\psi} \cdot \psi + 2g)\,\psi + i\, a\, \psi. \eqno (B.3) \]
Multiplication from the left by $\bar{z}$ in the above equation demonstrate that $\bar z \cdot {\dot \psi} = 0$ due to
the constraint equation $\bar z \cdot \psi = 0$.  This, in turn, implies that $\dot {\bar z} \cdot \psi = 0$ 
(due to the observation  that $d/dt\, (\bar z \cdot \psi) = 0 \Rightarrow \bar z \cdot \dot\psi + \dot{\bar z} \cdot \psi = 0$).
It   is now trivial to prove  that $\dot Q = 0$. We lay emphasis on the fact that it is an elegant combination 
of the equations of motion (10) and the constraints ($\bar z\cdot z = 1\,
\bar z \cdot \psi = 0,\, \bar\psi \cdot z = 0$) which are to be invoked for the proof of $\dot Q = 0$.
In exactly similar fashion, it can be seen that $\dot {\bar Q} = 0$ due to the equations of motion (10) 
and the constraints $\bar \psi \cdot z = 0, \dot{\bar \psi} \cdot z = 0, \bar \psi \cdot {\dot z} = 0$.
The proof of $\dot H \equiv {\dot Q}_\omega = 0$ is elementary at the {\it classical}  as well as 
the {\it quantum} levels  because the Poisson bracket and/or commutator of $H$ with itself is  zero.\\

\vskip .5cm

\noindent
{\bf{\large{~~~~~~~~~Appendix C: Symmetries and Algebraic Structure}}}\\

\vskip .5cm

\noindent
We exploit here the ideas of continuous symmetries and their generators 
to derive one of the simplest  form of $sl(1/1)$ algebra that is satisfied 
amongst the conserved charges $(Q, \bar{Q}, Q_\omega)$ of our theory. By exploiting 
the concept of generators, it is trivial to note that
\[s_1\,Q = i\,\lbrace Q, \, {Q}\rbrace = 0, \qquad s_2\,\bar{Q} = i\,\lbrace\bar{Q}, \,
\bar{Q}\rbrace = 0, \eqno (C.1) \]
because when we compute the l.h.s. of the above equation by exploiting the expression 
for $(Q, \bar{Q})$ from (8) and symmetry transformations from (3), we observe that
\[ s_1\, Q = -\, \frac{1}{2}\, (D_t \bar z \cdot \psi) (\bar z \cdot \psi), \qquad \qquad
s_2\,\bar{Q} = \, \frac{1}{2}\, (\bar \psi \cdot D_t  z) (\bar \psi \cdot z), \eqno (C.2) \]
which turn out to be {\it zero} on the constrained surface defined by the constraint conditions  
$\bar{z}\cdot \psi = 0$ and $\bar{\psi}\cdot z = 0$, respectively. In  similar fashion,
we compute  the l.h.s. of the following relationships:
\[s_1\,\bar{Q} = i\,\lbrace \bar{Q}, Q\rbrace = i\,H, \qquad s_2\,Q 
= i\,\lbrace Q, \bar{Q}\rbrace = i\,H, \eqno (C.3) \]
by exploiting the inputs available in (8) and (3) to obtain the following explicit expressions:
\[ s_2\,Q  =   2\, i\, D_t \bar z \cdot D_t  z + D_t \bar \psi \cdot \psi  + \frac{1}{2}\,
(\bar\psi \cdot D_t z)(\bar z \cdot \psi) \] 
\[ +  \, \frac{i}{4}\, (\bar \psi \cdot \psi + 2g)\, (\bar\psi \cdot \psi) 
- \frac{1}{2}(\bar \psi \cdot \psi + 2g)\,(\bar z \cdot D_t z),\]
\[ s_1\,\bar{Q}  =   2\, i\, D_t \bar z \cdot D_t z  
- \,( \bar \psi \cdot D_t \psi) -\frac{1}{2}\, (D_t \bar z \cdot \psi) (\bar \psi \cdot z) \] 
\[ + \, \frac{i}{4}\, (\bar \psi \cdot \psi + 2g)\, (\bar\psi \cdot \psi) 
+ \frac{1}{2}\,(D_t \bar z \cdot z)\, (\bar \psi \cdot \psi + 2\, g). \eqno (C.4) \]
Now using the definitions  of $D_t z, D_t \bar z$ and  constraints 
$\bar{\psi} \cdot z = 0, \bar z\cdot \psi = 0$ 
plus the equations of motion for $\psi$ and $\bar{\psi}$  from (10), we obtain the following:
\[ s_1\,\bar{Q}  =   2\, i\, D_t \bar z \cdot D_t z  
+ \frac{1}{2}\,(\dot{ \bar z} \cdot z + i\, a \, \bar z \cdot z)\, (\bar \psi \cdot \psi + 2\, g) 
 -  \frac{i}{4}\, (\bar \psi \cdot \psi + 2g)\, (\bar\psi \cdot \psi),~~~~~~~\]
 \[s_2\,Q  =   2\, i\, D_t \bar z \cdot D_t  z   
- \frac{1}{2}\,(\bar \psi \cdot \psi + 2g)\,({ \bar z} \cdot \dot z - i\, a  \bar z \cdot z) 
-  \, \frac{i}{4}\, (\bar \psi \cdot \psi + 2g)\, (\bar\psi \cdot \psi).  \eqno (C.5) \]
We further exploit the definition of $a$ and constraint $\bar z \cdot z = 1$ and  
$\frac{d}{dt}\,(\bar{z}\cdot z - 1) = 0$, in the above equation, to obtain the following
\[ s_1\,\bar{Q} = i\,\left[2\,  D_t \bar z \cdot D_t z 
 - \frac{1}{2}\, (\bar \psi \cdot \psi + 2g)\, (\bar\psi \cdot \psi)\right ] \equiv iH, \]
\[ s_2\,Q =   i\,\left[2\,  D_t \bar z \cdot D_t z  - \frac{1}{2}\, (\bar \psi \cdot \psi 
+ 2g)\, (\bar\psi \cdot \psi)\right ] \equiv iH.   \eqno (C.6) \]
To be specific, in the computation of $s_1\, \bar Q = i \,\{\bar Q, \, Q\} = i \,H$,
we have used the constraint conditions $\bar \psi \cdot z = 0, \bar z\cdot z = 1, 
\frac{d}{dt}\, (\bar z\cdot z - 1) = 0$ and 
the definitions of $a, D_t \psi, D_t \bar z$.  On the other hand, in the explicit composition of 
$s_2\,  Q = i\, \{ Q, \, \bar Q\} = i \,H$, we have exploited the constraint conditions 
$\bar z \cdot \psi = 0, \bar z\cdot z = 1, \frac{d}{dt}\, (\bar z\cdot z - 1) = 0$ 
and the definitions of $a, D_t \bar\psi, D_t   z$.
Thus, we note that the constraints as well as the equations of motion are to be 
exploited  judiciously  to prove that $s_1 \bar Q = s_2 Q = i\, H$ which, ultimately, 
implies that $\{Q, \bar Q\} = H$.\\

\vskip .5cm

\noindent
{\bf{\large{~~~~~~~~~~~~~~~~~Appendix D: On Discrete Symmetries }}}\\ 

\vskip .5cm

\noindent
In addition to the discrete symmetry transformations (11), we have the following useful discrete symmetry 
transformations for the Lagrangian (1), namely;
\[ z \rightarrow  \pm  i \,\bar z, \,\qquad \bar z \rightarrow   \mp\, i\,  z, \,\qquad
\psi \rightarrow  \pm \, i\, \bar \psi, \qquad
 \bar\psi \rightarrow  \pm\, i \, \psi, \]
\[ t 
\rightarrow -\, t, \;\;\;\qquad a \rightarrow + \,a,\;\; \;\qquad g \to g. ~~~~~~~~~~~~~~~~~~~~~~~~ \eqno (D.1) \]
The above symmetry transformations obey all the conditions that have been satisfied by (11).
Thus, these symmetries are as good as our transformations  in (11). In fact, it can
be readily checked that $* L = L, * H = H, * Q = - \bar Q, * \bar Q = Q$ are true under (D.1).

We dwell a bit on a couple of discrete symmetry transformations for the Lagrangian (1) that
are {\it not} acceptable to us because they do not comply with the strictures laid down by
the duality invariant theories [30]. These transformations, for instance,  are
\[ z \rightarrow  \pm   \,\bar z, \,\qquad \bar z \rightarrow   \pm\, \,  z, \,\qquad
\psi \rightarrow  \pm \, \, \bar \psi, \qquad
 \bar\psi \rightarrow  \pm\,  \, \psi, \] \[ t 
\rightarrow +\, t, \;\;\qquad a \rightarrow - \,a, \;\,\qquad g \to - g,~~~~~~~~~~~~~~~~~~~ \eqno (D.2) \]
which leave the Lagrangian (1) invariant. It can be checked that
 \[ * \,( *\, z ) = z, \qquad * \,( *\, \bar z ) = \bar z, \qquad * \,( *\, \psi ) = \psi, \qquad 
* \,( *\, \bar \psi ) =  \bar \psi. \eqno (D.3)\]
With the above  observations, we can verify that the following is true:
\[ s_2 \; \Phi = + \, *\,s_1\, *\, \Phi, \qquad \qquad \Phi = z, \bar z, \psi, \bar\psi. \eqno (D.4)\]
However, we note that the reciprocal relation
 \[ s_1 \; \Phi = - \, *\,s_2\, *\, \Phi, \qquad \qquad \Phi = z, \bar z, \psi, \bar\psi, \eqno (D.5) \]
is {\it not} satisfied at all.  Thus, the discrete symmetry transformations (D.2) of the Lagrangian (1)
are  {\it not} acceptable because they do not comply with the conditions (e.g. reciprocal relationship) 
laid down by the duality invariant theories  [30]. \\

\end{document}